# First-principles study of the bandgap renormalization and optical property of β-LiGaO$_2$

Dangqi Fang[1]*

[1]MOE Key Laboratory for Nonequilibrium Synthesis and Modulation of Condensed Matter, School of Physics, Xi'an Jiaotong University, Xi'an 710049, China

Abstract

β-LiGaO$_2$ with an orthorhombic wurtzite-derived structure is a candidate ultrawide direct-bandgap semiconductor. In this work, using the non-adiabatic Allen-Heine-Cardona approach, we investigate the bandgap renormalization arising from electron-phonon coupling. We find a sizable zero-point motion correction of -0.362 eV to the gap at Γ, which is dominated by the contributions of long-wavelength longitudinal optical phonons. The bandgap of β-LiGaO$_2$ decreases monotonically with increasing temperature. We investigate the optical spectra by comparing the model Bethe-Salpether equation method with the independent-particle approximation. The calculated optical spectra including electron-hole interactions exhibit strong excitonic effects, in qualitative agreement with experiment. The contributing interband transitions and the binding energy for the excitonic states are analyzed.

*fangdqphy@xjtu.edu.cn

# I. Introduction

LiGaO$_2$ has been of interest for use as a promising lattice-matched substrate for the epitaxial growth of nonpolar GaN [1] and ZnO films [2]. The ground-state structure of LiGaO$_2$ is the orthorhombic Pna2$_1$ structure, referred to as β-LiGaO$_2$, for which the prototype is β-NaFeO$_2$. This material is a ternary analogue of wurtzite ZnO, which is arrived at by replacing the Zn$^{2+}$ ions alternatively by the Li$^+$ and Ga$^{3+}$ ions. The LiGaO$_2$ crystal can be grown by the Czochralski method [3], and atomically flat LiGaO$_2$ surfaces were reported [4]. Recently, researches on LiGaO$_2$ have also focused on possible application in optoelectronic devices. LiGaO$_2$ can be alloyed with ZnO by a standard solid state reaction, forming $x$(LiGaO$_2$)$_{1/2}$-(1-$x$)ZnO ($x \leq 0.38$) semiconductor alloys with tunable bandgap between 3.33 eV and 3.7 eV [5]. Solid solutions of β-CuGaO$_2$ and β-LiGaO$_2$, i.e., β-(Cu$_{1-x}$Li$_x$)GaO$_2$, were synthesized, for which the bandgap changes from 1.47 eV to 3 eV in the composition range of $0 \leq x \leq 0.89$, covering the full visible range [6]. Defect studies show that the equilibrium carrier concentrations induced by the native point defects of LiGaO$_2$ are negligible [7], while n-type doping should be readily achieved with Si or Ge dopants, making LiGaO$_2$ a candidate ultrawide bandgap semiconductor [8].

The bandgap of β-LiGaO$_2$ has been investigated experimentally and theoretically in the literature [9–14]. A bandgap of 5.6 eV was derived from the soft x-ray spectroscopy analysis [9], while Ref. [10] gave a direct gap of 5.26 eV by the optical absorption measurement. The theoretical bandgap of β-LiGaO$_2$ calculated with density functional theory (DFT) method depends strongly on the exchange-correlation functional used.

Both local and semi-local functionals severely underestimate the bandgap of $\beta$-LiGaO$_2$, e.g. 3.136 eV obtained with the local density approximation (LDA) [13] and 3.363 eV with the Perdew-Burke-Ernzerhof (PBE) generalized gradient approximation [14]. Johnson *et al.* using the modified Becke-Johnson functional (mBJ) predicted a direct gap of 5.658 eV, in reasonable agreement with the soft x-ray spectroscopy measurements [9]. A recent study using the quasiparticle self-consistent *GW* (QS*GW*) method in the 0.8$\Sigma$ approximation reported a gap of 5.81 eV based on the experimental lattice constants of $\beta$-LiGaO$_2$ [14], and this study also estimated a zero-point motion correction of -0.2 eV to the bandgap by the electron-phonon coupling using a model with a single longitudinal optical (LO) phonon. Taking the zero-point renormalization (ZPR) into account, Ref. [14] gave a gap of 5.6 eV, in excellent agreement with the experimental value in Ref. [9], manifesting the importance of the electron-phonon interaction in this material.

The influence of electron-phonon coupling on the electronic structure of semiconductors and insulators is nonnegligible in many cases and has attracted considerable attention recently [15–20]. However, the first-principles calculation of the bandgap renormalization in polar materials is difficult. The LO phonons at long wavelength generate macroscopic electric fields, and the electron-phonon vertex diverges for $\mathbf{q} \to 0$, where $\mathbf{q}$ is the phonon wave vector. To address this challenge, Verdi *et al.* proposed a strategy that separates the short-range and the long-range contributions to the electron-phonon matrix elements, which generalizes the Fröhlich theory and can be used in conjunction with *ab initio* interpolation method for efficient and accurate calculation [21]. Miglio *et al*. [18] highlighted the importance of non-adiabatic effects in the ZPR of the electronic band gap and showed that for the materials with light elements, the magnitude of ZPR is often larger than 0.3 eV, and up to 0.7 eV, which cannot be ignored for accurate bandgap calculation in these materials.

In this work, we investigate the bandgap renormalization of β-LiGaO$_2$ induced by the electron-phonon coupling using first-principles calculations, finding a sizable zero-point motion correction of -0.362 eV to the bandgap. Furthermore, to explore the optical spectra, we perform separate calculations including the electron-hole interaction based on the equilibrium static crystal structure.

## II. Computational details

A. Electronic band structure

Our first-principles calculations for the structural relaxation and electronic band structure are based on DFT as implemented in the Vienna *ab initio* simulation package (VASP) [22]. The electron-ion interaction is described by means of the projector augmented-wave method [23,24] and the PBEsol exchange-correlation functional is used [25]. The cut-off energy for the plane wave basis set is set to 550 eV and a 4×4×4 Monkhorst-Pack **k**-point grid is employed for the Brillouin zone (BZ) integration [26]. All atoms are relaxed until the Hellmann-Feynman forces on them are less than 0.01 eV/Å. For the quasiparticle $G_0W_0$ and $GW_0$ investigations, the PBEsol equilibrium structure, 96 frequency grid points, and 1200 bands are used, with a cut-off energy of 200 eV for the response function [27].

B. Electron-phonon coupling

The effect of electron-phonon coupling on the electronic band structure is studied with the ABINIT (version 9.6.2) code [28,29] using the PseudoDojo PBEsol norm-conserving pseudopotentials [30], and setting the plane wave cut-off energy to 1088.46 eV. A 4×4×4 Γ-centered **k**-point grid for the BZ sampling is employed to obtain the electronic ground state density. For the phonon calculations, a coarse 4×4×4 Γ-centered **q**-point grid is used with the density functional perturbation theory (DFPT) approach, which gives converged phonon dispersion.

In the many-body perturbation theory approach, the electron-phonon self-energy $\Sigma_{n\mathbf{k}}^{e-ph}$ consists of two terms, namely the Fan-Migdal (FM) and Debye-Waller (DW) terms, at the lowest order of perturbation [31,32]:

$$\Sigma_{n\mathbf{k}}^{e-ph}(\omega,T) = \Sigma_{n\mathbf{k}}^{FM}(\omega,T) + \Sigma_{n\mathbf{k}}^{DW}(T). \tag{1}$$

The frequency-dependent FM term is given as

$$\Sigma_{n\mathbf{k}}^{FM}(\omega,T) = \sum_{m,\nu} \int_{BZ} \frac{d\mathbf{q}}{\Omega_{BZ}} |g_{mn\nu}(\mathbf{k},\mathbf{q})|^2$$
$$\times \left[ \frac{n_{\mathbf{q}\nu}(T) + f_{m\mathbf{k}+\mathbf{q}}(T)}{\omega - \varepsilon_{m\mathbf{k}+\mathbf{q}} + \omega_{\mathbf{q}\nu} + i\eta} + \frac{n_{\mathbf{q}\nu}(T) + 1 - f_{m\mathbf{k}+\mathbf{q}}(T)}{\omega - \varepsilon_{m\mathbf{k}+\mathbf{q}} - \omega_{\mathbf{q}\nu} + i\eta} \right], \tag{2}$$

where the phonon modes are denoted by indices $\nu$, wave vector **q**, and energy $\omega_{\mathbf{q}\nu}$, and $f_{m\mathbf{k}+\mathbf{q}}(T)$ and $n_{\mathbf{q}\nu}(T)$ are the Fermi-Dirac and Bose-Einstein distribution functions, respectively. The electron-phonon matrix elements $g_{mn\nu}(\mathbf{k},\mathbf{q})$ are defined by

$$g_{mn\nu}(\mathbf{k},\mathbf{q}) = \langle \psi_{m\mathbf{k}+\mathbf{q}} | \Delta_{\mathbf{q}\nu} V^{KS} | \psi_{n\mathbf{k}} \rangle, \tag{3}$$

with $\Delta_{\mathbf{q}\nu}V^{KS}$ the first-order variation of the self-consistent Kohn-Sham (KS) potential induced by the phonon mode $\mathbf{q}\nu$. The static DW term is approximated with [32]

$$\Sigma_{n\mathbf{k}}^{DW}(T) = \sum_{\mathbf{q}\nu m}\left[2n_{\mathbf{q}\nu}(T)+1\right]\frac{g_{mn\nu}^{2,DW}(\mathbf{k},\mathbf{q})}{\varepsilon_{n\mathbf{k}}-\varepsilon_{m\mathbf{k}}}, \quad (4)$$

where $g_{mn\nu}^{2,DW}(\mathbf{k},\mathbf{q})$ is an effective matrix element that, within the rigid-ion approximation, can be expressed in terms of the $g_{mn\nu}(\mathbf{k},\mathbf{q})$ matrix elements.

We use a Fourier-based interpolation technique to obtain the phonon coupling potentials on arbitrary **q**-points [28,33,34]. Meanwhile, a special numerical treatment is used to address the long-range behavior of the phonon coupling potential in polar semiconductors. As the discussion in Refs. [21,28,34], the long-range part associated to a displacement of atom $\kappa$ along the cartesian direction $\alpha$ is defined as

$$V_{\kappa\alpha\mathbf{q}}^{\mathcal{L}}(\mathbf{r}) = i\frac{4\pi}{\Omega}\sum_{\mathbf{G}\neq -\mathbf{q}}\frac{(\mathbf{q}+\mathbf{G})_{\beta}\cdot \mathbf{Z}_{\kappa\beta,\alpha}^{*}e^{i(\mathbf{q}+\mathbf{G})\cdot(\mathbf{r}-\boldsymbol{\tau}_{\kappa})}}{(\mathbf{q}+\mathbf{G})\cdot\boldsymbol{\varepsilon}^{\infty}\cdot(\mathbf{q}+\mathbf{G})}, \quad (5)$$

where $\mathbf{Z}^{*}$ and $\boldsymbol{\varepsilon}^{\infty}$ are the Born effective charge tensor and the dielectric tensor, respectively, and summation over the cartesian directions $\beta$ is implied. The long-range part is subtracted from the phonon coupling potential to yield the short-range part of the potential. The short-range part of the phonon coupling potential is then Fourier transformed from the coarse **q**-point mesh to a real space lattice-vector mesh. The interpolated short-range part of the phonon coupling potential for an arbitrary **q** point can be obtained via the inverse Fourier transformation. The long-range part is added to recover the full phonon coupling potential. This scheme reproduces accurately the electron-phonon coupling matrix elements [34].

We employ the non-adiabatic Allen-Heine-Cardona (AHC) approach [18] to evaluate zero-point and temperature-dependent renormalizations of the electronic band structure. The correction to the energy of the electronic state $n\mathbf{k}$ is obtained from the real part of the self-energy, Eq. (1), evaluated at the bare KS eigenvalue, i.e. $\omega = \varepsilon_{n\mathbf{k}}$: [31,32]

$$\Delta\varepsilon_{n\mathbf{k}}(T) = \text{Re}\left[\Sigma_{n\mathbf{k}}^{e-ph}(\varepsilon_{n\mathbf{k}},T)\right]. \tag{6}$$

C. Optical property

We use two methods to calculate the optical spectrum. The first one is based on the independent-particle picture. The frequency-dependent dielectric matrix in the long-wavelength limit is calculated using VASP code [35] and a 10×10×10 Γ-centered **k**-point grid. A scissors correction [36,37] is used in order to reproduce the $GW_0$+ZPR gap.

Beyond the independent-particle approximation, the electron-hole interactions may significantly alter the optical spectrum. These can be accounted for by solving the Bethe-Salpether equation (BSE) [38,39]. However, calculations on dense k-point grids at the standard $GW$+BSE level for the 16-atom primitive cell of β-LiGaO$_2$ are prohibitive due to the huge computational cost. Alternatively, we employ an analytic model for the static screening. In this model BSE (mBSE) the dielectric function is described as [40]

$$\varepsilon_{\mathbf{G},\mathbf{G}}^{-1}(\mathbf{k}) = 1 - \left(1 - \varepsilon_\infty^{-1}\right)\exp\left(-\frac{|\mathbf{k}+\mathbf{G}|^2}{4\lambda^2}\right), \tag{7}$$

where $\varepsilon_\infty$ and $\lambda$ are the ion-clamped static dielectric function and the range-separation parameter, respectively, which are determined by fitting $\varepsilon_{\mathbf{G},\mathbf{G}}^{-1}(\mathbf{k})$ to the screened Coulomb kernel diagonal values obtained from the $GW_0$ calculation. The off-diagonal elements of the inverse dielectric functional are neglected making the screened Coulomb kernel diagonal ($\mathbf{G} = \mathbf{G}'$). The inverse of the dielectric function from $GW_0$ and the fit according to Eq. (7) are shown in Fig. S3 (see Supplementary material). The parameters, obtained from the fit to the $GW_0$ dielectric function, $\lambda$ =1.3981 and $\varepsilon_\infty^{-1}$ =0.3147, are used as input for the mBSE calculations. The quasiparticle energies are approximated by the application of the scissors operator with a shift corresponding to the difference between the $GW_0$+ZPR gap and the PBEsol gap to the PBEsol single-particle energies. The Tamm-Dancoff approximation, 10 occupied and 10 unoccupied bands, and 10×10×10 Γ-centered **k**-point grid are used for the mBSE calculations.

### III. Results and discussion

A. Structural and electronic properties

We first optimize the lattice constants of β-LiGaO$_2$ using the 16-atom primitive cell. The values obtained with the PBEsol functional are $a$=5.399 Å, $b$=6.367 Å, and $c$=5.015 Å, as shown in Table I, which are closer to the experimental data than the previous results obtained using the LDA or PBE functionals. Figure 1 shows the calculated electronic band structure of β-LiGaO$_2$, which gives a direct bandgap of 3.261 eV across the Γ point. The valence bands between -6 eV and 0 eV are dominated by the O 2$p$ states, and the narrow bands at around -12 eV are the Ga 3$d$ derived bands. The bottom of conduction band is mainly contributed by the Ga 4$s$ state. These features are in good agreement with prior studies [9,14]. In Fig. 2, we zoom in on the region near the valence band maximum (VBM) and the conduction band minimum (CBM) and present the orbital contribution for each band. The top three valence bands near the Γ point are formed by the $p_z$, $p_x$, and $p_y$ orbitals due to the crystal-field splitting. The VBM, VBM-1, VBM-2 states labeled according to the irreducible representations a1, b1, and b2 of the point group $C_{2v}$ have the same symmetry as a vector along $z$, $x$, and $y$, respectively, chosen along the $c$, $a$, and $b$ lattice axes. The CBM state corresponds to the a1 representation. The symmetry of the electronic states plays an important role in determining whether the optical transition between them can occur; for example, the optical transition from the a1 VBM state to the CBM at the Γ point are dipole-allowed only for **E** ∥ ***c***, from the b1 VBM-1 state to the CBM for **E** ∥ ***a***, and from the b2 VBM-2 state to the CBM for **E** ∥ ***b***.

The effective masses of electron and hole along the $x$, $y$, and $z$ directions obtained using the VASPKIT code [41] are given in Table II. The electron effective mass is about 0.4 (in unit of free electron mass), almost independent of the directions. The hole in each valence band has one light mass in the direction corresponding to the symmetry of the band and two heavy masses in two other directions. Compared with the values obtained in Ref. [14] using the 0.8∑-QS*GW* method, our calculated electron masses agree well, while some heavy hole masses have small discrepancy.

To overcome the bandgap underestimation problem of the semilocal functional, we perform quasiparticle $G_0W_0$ and $GW_0$ calculations. The bandgap at Γ is calculated to be 5.538 eV and 5.995 eV for the $G_0W_0$ and $GW_0$, respectively. The $GW_0$ gap is larger than the experimental values, 5.6 eV from Ref. [9] and 5.26 eV from Ref. [10], but comparable with the 0.8∑-QS*GW* gap of 5.81 eV [14]. The reason for the difference from the experiments may be that our calculation does not account for the electron-phonon coupling effect for the electronic bandgap.

B. Bandgap renormalization due to the electron-phonon coupling

In Fig. S1, we test the convergence of the ZPR of the bandgap of β-LiGaO$_2$ in terms of **q**-point sampling. We find the ZPR obtained with a 20×20×20 **q**-point grid is -0.354 eV, while that with a 24×24×24 **q**-point grid is -0.362 eV, showing that we have reached convergence to 0.01 eV. It is interesting that this ternary oxide has a large ZPR. Using the first-principles non-adiabatic AHC methodology Miglio *et al*. [18] found that the ZPR for ZnO is -0.157 eV (the experimental data is -0.164 eV [42]) and for MgO -0.524 eV. β-LiGaO$_2$ lies between ZnO and MgO in terms of the value of ZPR. Taking electron-phonon coupling into account, the $GW_0$+ZPR gap is 5.633 eV, as shown in Table III, that agrees very well with the experimental result of 5.6 eV from the soft x-ray spectroscopy measurements [9].

Figure 3 (a) show the temperature dependence of the bandgap renormalization. The magnitude of the bandgap renormalization increases monotonically with elevating temperature. The absolute value of bandgap renormalization changes by 0.126 eV with temperature increasing from 0 K to 300 K. In Fig. 3 (b), we present the vibrational correction to the VBM and CBM. The VBM correction is 0.228 eV and the CBM correction -0.134 eV due to zero-point motion. From the Fröhlich model [18,43], the CBM correction can be expressed as $\text{ZPR}_c = -\left(\dfrac{1}{\varepsilon_\infty} - \dfrac{1}{\varepsilon}\right)\sqrt{\dfrac{m^*\omega_{LO}}{2}}$. The VBM correction has similar expression but opposite sign. The smaller correction for CBM is mainly due to the smaller effective mass for electrons. Bhosale *et al.* [44] showed that materials with $d$ electrons in the valence band (e.g., $CuGaS_2$, $AgGaS_2$) exhibit a nonmonotonic temperature dependence of the energy gap associated with low-energy vibrations of the $d$-electron elements, while those without $d$ electrons (e.g., $ZnSnAs_2$) exhibit the standard gap decrease with increasing temperature. For $\beta$-$LiGaO_2$, the valence bands between -6 eV and 0 eV have no $d$ electrons and its behavior agrees with the result of Ref. [44].

To explore the origin of the bandgap renormalization, we estimate the ZPR using the generalized Fröhlich (gFr) model proposed in Ref. [18]. In this model, the CBM correction $ZPR_c^{gFr}$ is given by

$$ZPR_c^{gFr} = -\sum_{jn} \frac{1}{\sqrt{2}\Omega_0 n_{deg}} \int_{4\pi} d\mathbf{q} \left(m_n^*(\mathbf{q})\right)^{1/2} \times \left(\omega_{j0}(\mathbf{q})\right)^{-3/2} \left(\frac{\mathbf{q} \cdot \mathbf{p}_j(\mathbf{q})}{\epsilon^\infty(\mathbf{q})}\right)^2, \quad (8)$$

where $\Omega_0$ and $n_{deg}$ represent the primitive cell volume and the degeneracy of the band edge, respectively, $m_n^*(\mathbf{q})$ is the electron effective mass, $\omega_{j0}(\mathbf{q})$ denotes the $\mathbf{q} \to 0$ limit of the $j$-phonon branch, i.e. $\omega_{j0}(\mathbf{q}) = \lim_{q \to 0} \omega_{j(q\mathbf{q})}$, and $\mathbf{p}_j(\mathbf{q})$ is the mode-polarity vector. A similar expression exists for the VBM correction. This model includes the features of multiphonon, anisotropic, degenerate band extrema for real material and can be readily evaluated only at $\mathbf{q} = \Gamma$. The ZPR for $\beta$-$LiGaO_2$ from the generalized Fröhlich model is found to be -0.307 eV, which captures 85% of the first-principles result, demonstrating that the large ZPR of $\beta$-$LiGaO_2$ is dominated by the contributions from the long-wavelength LO phonon modes.

C. Optical property

Next, we discuss the optical spectrum of β-LiGaO$_2$. The calculated optical spectra based on the equilibrium static crystal structure are illustrated in Fig. 4 (a). The spectra obtained within the independent-particle approximation show different onsets for different polarizations, consistent with the symmetry analysis for the electric dipole transition presented above; for example, the onset of the spectrum is at 5.63 eV for **E** || **c**, 5.71 eV for **E** || **a**, and 5.74 eV for **E** || **b**, respectively. This trend agrees well with the study of Radha *et al*. that used the 0.8∑-QS*GW* method [14].

The inclusion of the mBSE correction shows strong excitonic contributions to the spectrum. The onsets of the spectra of different polarizations are redshifted dramatically with pronounced exciton absorption peaks at 5.20 eV (**E** || **c**), 5.27 eV (**E** || **a**), and 5.32 eV (**E** || **b**), respectively. The characteristics of our calculated spectra are in reasonable agreement with the experimental result of Tumėnas *et al*. [11], though in the experiment the positions of the peaks are higher in energy overall. The exciton wave function is expressed in the electron-hole product basis [40], $\Phi^i = \sum_{cv\mathbf{k}} A^i_{c,v,\mathbf{k}} \phi_{c,\mathbf{k}} \phi_{v,\mathbf{k}}$. In Figs. 4(b)-4(d), we plot the fat-band pictures according to the coefficients $|A^i_{c,v,\mathbf{k}}|$ of the exciton wave functions for the three marked transitions. The transition #1 is localized at Γ and is between the O 2$p_z$ states at the VBM and the CBM states. The transition #2 (#3) is also localized at Γ, but involve the O 2$p_x$ (2$p_y$) states and the CBM states.

Form the data above, we can get an average binding energy of 0.43 eV for the excitons corresponding to the transitions #1, #2, and #3. We note that for MgO, a wide gap semiconductor with a gap of about 7.8 eV [45], the exciton binding energy is calculated to be 0.429 eV by Fuchs *et al*. [46], 0.37 eV by Filip *et al*. [47], while the experimental value is 0.08 eV [45] and 0.145 eV [48]. This discrepancy is attributed to the neglect of polaronic [40] and phonon screening effects [47]. Our mBSE calculations for β-LiGaO$_2$ do not involve the polaronic and phonon effects, likely leading to an overestimated exciton binding energy.

We estimate the binding energy E$_{xb}$ of the lowest exciton state using a simple hydrogen-like atom model, i.e. $E_{xb} = 13.606 \mu/(m_e \varepsilon^2)$ (in eV) with $\mu$ the effective mass, m$_e$ the free electron mass, and $\varepsilon$ the dielectric constant [49]. Using an average hole effective mass of about 3 m$_e$ and an electron effective mass of 0.4 m$_e$, we obtain $\mu$ = 0.35 m$_e$. We choose the static dielectric constant (including ionic contribution) of 7 for $\varepsilon$, finding a reduced E$_{xb}$ of 0.097 eV, in line with the rough estimate of about 50 meV from the experiment of Tumėnas *et al*. [11].

Recently, studies of polarons from first principles have advanced considerably. A secular equation involving phonons and electron-phonon matrix elements was proposed to investigate polaron energies and wave functions, without resorting supercell calculations [50,51]. A self-consistent many-body theory of electron-phonon couplings was developed, which unifies the calculations of phonon-induced band structure renormalization and polaron localization [52,53]. The localization effects on the band structure renormalization of β-LiGaO$_2$ is worth investigating in future work. Moreover, recent developments of polaron calculations [50–53] make possible the exploration of polaronic effects on the exciton property.

## IV. Conclusions

In this work, using the non-adiabatic Allen-Heine-Cardona approach, we find a sizable ZPR of -0.362 eV for β-LiGaO$_2$, rendering the $GW_0$ +ZPR gap in good agreement with the experimental result. The generalized Fröhlich model reveals that the ZPR is dominated by the contributions from the long-wavelength LO phonon modes. The magnitude of the bandgap renormalization increases monotonically with increasing temperature. The optical spectra exhibit strong excitonic effects, in qualitative agreement with experiment. The exciton binding energy tends to be overestimated by the first-principles calculations with the lack of polaronic and phonon screening effects. Our study is helpful to deeply understand the electronic structure and optical property of β-LiGaO$_2$ and facilitate its application in optoelectronic devices.

**Supplementary Material**

See supplementary material for the convergence tests, phonon dispersion of β-LiGaO$_2$, and the fit for the input parameters of the mBSE.

**ACKNOWLEDGMENTS**

We acknowledge the financial support from the National Natural Science Foundation of China (Grant No. 11604254) and the Natural Science Foundation of Shaanxi Province (Grant No. 2019JQ-240). We also acknowledge the HPCC Platform of Xi'an Jiaotong University for providing the computing facilities.

TABLE I. Calculated and experimental lattice constants (in Å) of β-LiGaO$_2$.

|  | Method | a | b | c |
| --- | --- | --- | --- | --- |
| This work (VASP) | PBEsol | 5.399 | 6.367 | 5.015 |
| This work (ABINIT) | PBEsol | 5.397 | 6.361 | 5.011 |
| Ref. [13] | LDA | 5.361 | 6.255 | 4.953 |
| Ref. [14] | PBE | 5.4665 | 6.4570 | 5.094 |
| Expt. [54] |  | 5.402 | 6.372 | 5.007 |

TABLE II. Electron and hole effective masses (in units of free electron mass) of β-LiGaO$_2$ and energy levels at Γ relative to the VBM calculated using the PBEsol functional within the VASP code. The values in parentheses correspond to the 0.8∑-QS*GW* result in Ref. [14].

| Band | E (eV) | m$_x$ | m$_y$ | m$_z$ |
| --- | --- | --- | --- | --- |
| CBM | 3.261 | 0.39 (0.39) | 0.42 (0.39) | 0.38 (0.41) |
| VBM | 0 | 4.37 (3.85) | 3.57 (3.50) | 0.43 (0.42) |
| VBM-1 | -0.077 | 0.46 (0.45) | 4.52 (3.50) | 4.83 (3.80) |
| VBM-2 | -0.109 | 4.65 (3.15) | 0.56 (0.58) | 4.04 (3.80) |

TABLE III. Bandgap (in eV) of β-LiGaO$_2$ from this work and other references.

| This work | Bandgap at Γ |
|---|---|
| PBEsol (VASP) | 3.261 |
| PBEsol (ABINIT) | 3.300 |
| $G_0W_0$ | 5.538 |
| $GW_0$ | 5.995 |
| $GW_0$+ZPR | 5.633 |
| Previous work | |
| LDA [13] | 3.136 |
| PBE [14] | 3.363 |
| mBJ [9] | 5.658 |
| 0.8∑-QS$GW$ [14] | 5.81 |
| Experiment [9] | 5.6 |
| Experiment [10] | 5.26 |

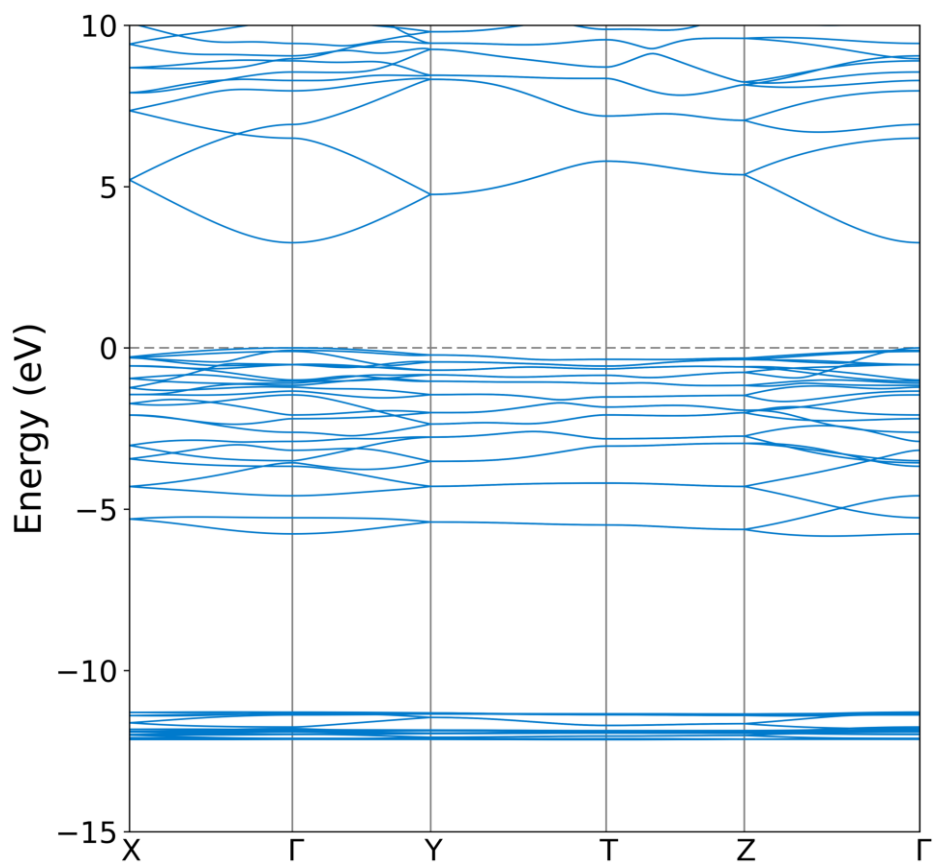

Fig. 1 Electronic band structure of β-LiGaO$_2$ (space group Pna2$_1$) calculated using the PBEsol functional. The valence band maximum is set to zero.

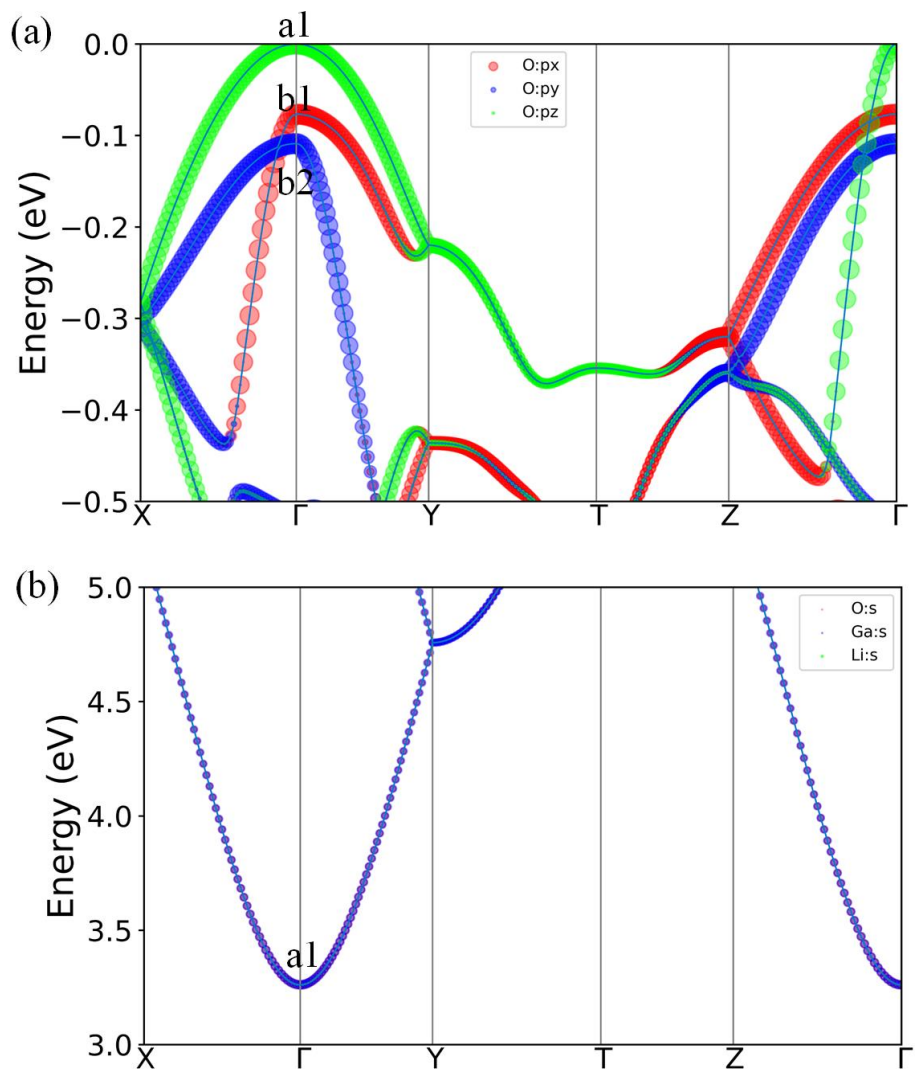

Fig. 2 Orbital-decomposed band structure for the top of valence band (a) and the bottom of conduction band (b). The Γ-point states are labeled according to the irreducible representations of point group $C_{2v}$.

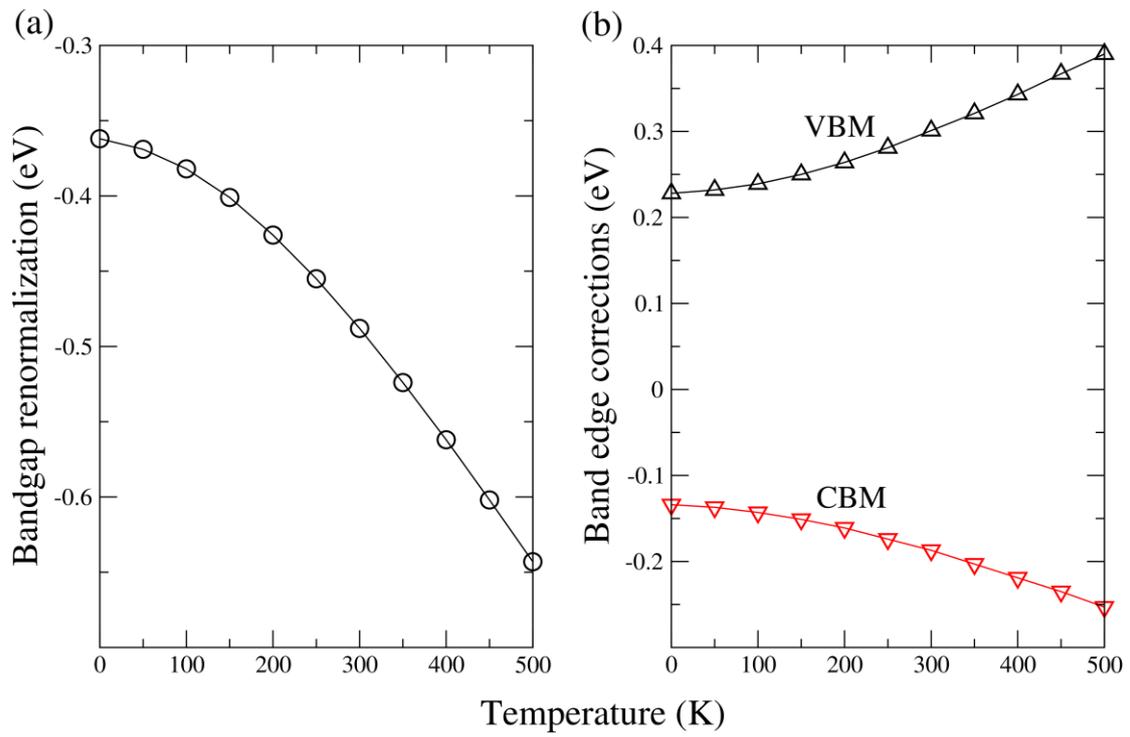

Fig. 3 Temperature dependence of the bandgap renormalization (a) and the VBM and CBM corrections (b) for β-LiGaO$_2$ due to electron-phonon coupling obtained with a 24×24×24 **q**-point grid.

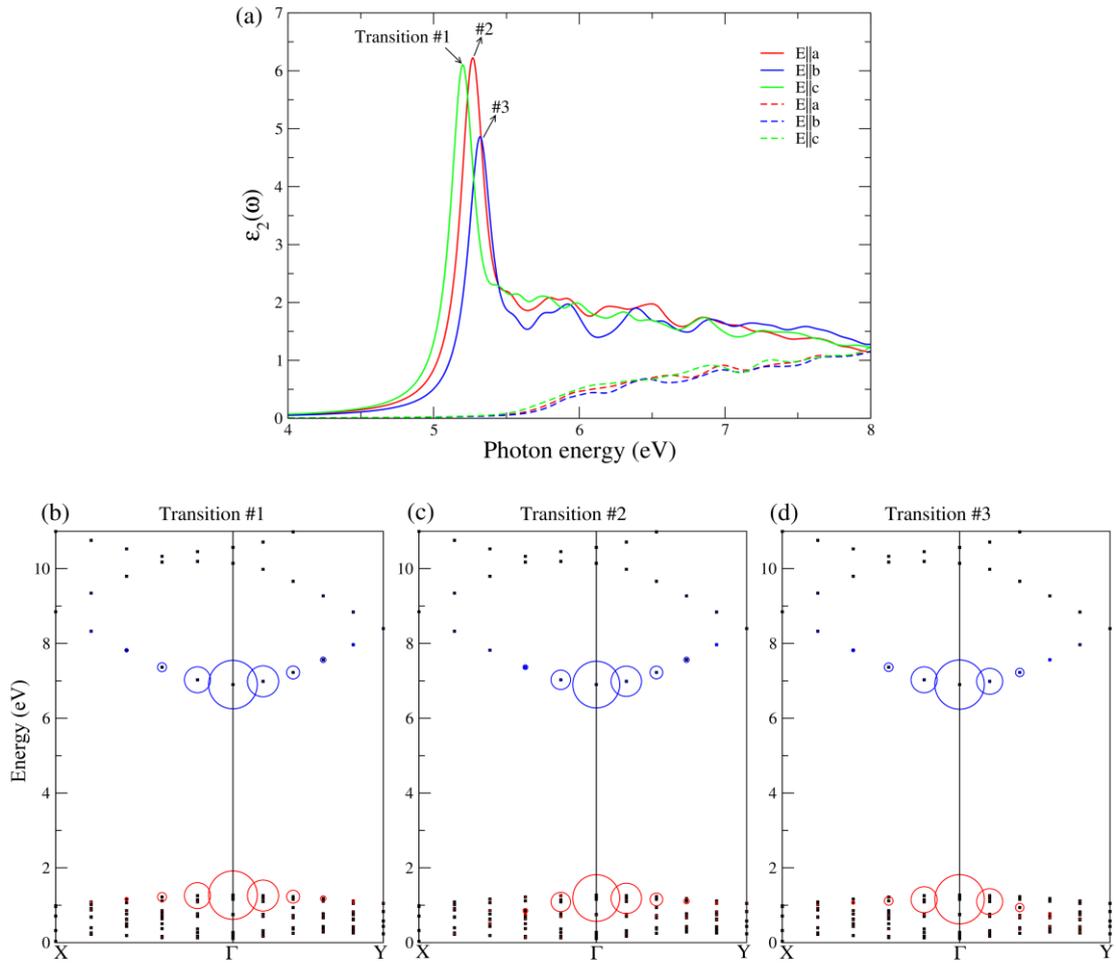

Fig. 4 (a) Imaginary part of the dielectric function of β-LiGaO$_2$ for light polarized along the *a*, *b*, and *c* directions. The spectra correspond to the ones obtained using the mBSE with 10 occupied band and 10 unoccupied band (solid lines), and to those within the long-wavelength limit and independent-particle approximation (dashed lines). (b)-(d) Fat-band pictures of the marked transitions.